\documentstyle[12pt]{article}
\begin{document}

\vspace{2.5cm}

\begin{center}
{\Large \bf
$SU(2)$ gauge theory in the maximally abelian gauge without monopoles.}\\
\vspace{1cm}
{\large S.Yu.~Shmakov and \underline{A.M.~Zadorozhny}
\footnote{email: zador@lcta32.jinr.dubna.su}}\\
{\large \it
Laboratory of Computing Techniques and Automation,\\ Joint Institute for
Nuclear Research, 141980 Dubna, Russia.}\\

\vspace{.5cm}

\begin{abstract}
We present an algorithm for simulation of $SU(2)$ lattice gauge theory under
the MA gauge and first numerical results for the theory without abelian
monopoles. The results support the idea that nonperturbative interaction arises
between monopoles and residual abelian field and the other interactions are
perturbative. It is shown that the Gribov region for the theory with the MA
gauge fixed is non-connected.
\end{abstract}
\end{center}

\section{Introduction}

In the last ten years there have been extensive studies of abelian monopole
dynamics inspired by conjectured dual-superconductor confinement mechanism
\cite {tHooft1, tHooft2}. The maximaly abelian (MA) gauge projection
\cite{Kron1} was found having abelian dominance and monopole condensation
\cite{BrndHioki} in $SU(2)$ gauge theory. It suggests the existence of an
effective $U(1)$ theory of confinement.

But questions about role of Gribov's copies \cite{HBMMP} and lattice artifacts
are not clear yet. Calculation of monopole condensate like \cite{Giac} and
direct investigation of the model without monopoles are also interesting. For
these reasons a simulation of $SU(2)$ gauge theory under the MA gauge
constraints is desirable.

This is the purpose of this work to present an algorithm for simulation of
$SU(2)$ gauge theory under the MA projection and first results of studying
topology of the field manifold. Four-link analysis of the effective action
for MA gauge is done in \cite{CPV}. We will keep here full effective action
for the case of $SU(2)$ lattice gluo-dynamics.

In the next section we briefly discuss the Faddeev--Popov operator for the MA
projection of $SU(2)$ lattice gauge theory and a partial solution of the
gauge constraints. We also give there a short description of a hybrid Monte
Carlo algorithm of the simulation \cite{Forthcome}. The numerical results
and concluding remarks are presented in the last two sections.

\section{$SU(2)$ under the MA gauge.}

The MA projection for a lattice gauge field configuration $\{ \tilde {U} \}$
in $SU(2)$ theory is defined by \cite{Kron1}

$$
\max_{\{g\}} R(\{ \tilde{U} \}, \{ g \})=
\max_{\{g\}} \sum_{l=(i,~j)} Tr[\sigma_3 U_l \sigma_3 U_l^+]
$$

\noindent
where $U_l = g_i \tilde{U_l} g_j^+$ are gauge transformed link variables and
$g_i,~g_j$ are arbitrary $SU(2)$ matrixes of the gauge transformation at
sites $i$ and $j = i+\mu$. Written as a stationarity condition it reads

\begin{equation}
\left. \frac {\delta R} {\delta \tilde{g_i}} \right|_{\tilde{g}=1} =
\sum_{\mu} Tr[\sigma_3 (U_{i+\mu} \sigma_3 U_{i+\mu}^+
+ U_{i-\mu}^+ \sigma_3 U_{i-\mu})] = Tr (\sigma_3 X_i) = 0
\label{mag}
\end{equation}

\noindent
or equivalently $$ X_i^{\bot} = 0 $$
\noindent
where $\tilde{g}_i$ is an $SU(2)$ matrix with $\tilde{g}_{3,i}$ taken to be
zero.  $X = \sigma_1 X_1 + \sigma_2 X_2 + \sigma_3 X_3$ is a traceless $2
\times 2$ antihermitian matrix and $X^{\bot} = \sigma_1 X_1 + \sigma_2 X_2$.

Partition function for the theory with gauge conditions (\ref{mag}) could be
written in the form

\begin{equation}
Z=\int \delta(X^\bot)\ det(F) \  \left[ d\mu (U) \right] \exp (- S_W\{U\})
\label{zma}
\end{equation}

\noindent
where $d \mu (U)$ is the invariant integration measure, $F$ is the
Faddeev-Popov operator for the MA gauge and $S_W$ is the Wilson action.

According to Gribov's idea \cite{Gribov} stationarity eqs.(\ref{mag}) should
be complemented with the stability condition that the Faddeev--Popov operator
be positively defined $F>0$.

On the lattice
$F = f_{ij}=-\frac{\delta^2 R}{\delta g_i \delta g_j}$
is a square symmetric matrix with nonzero diagonal elements ($i=j$) and elements
for sites $i$ and $j$ connected by a link. On surface (\ref{mag}) $f_{ij}$
reads

\begin{equation}
f_{ij}  =  \left\{
 \begin{array} [c]{cr}

\left|
\begin{array}{cc}

X_{3,i} & 0 \\
0 & X_{3,i}
\end{array} \right| & \mbox{for } i = j \\

 & \\

\left|
\begin{array}{cc}
Tr(\sigma_2U_{ij}\sigma_2U_{ij}^+) & Tr(\sigma_2U_{ij}\sigma_1U_{ij}^+) \\
 & \\
Tr(\sigma_1U_{ij}\sigma_2U_{ij}^+) & Tr(\sigma_1U_{ij}\sigma_1U_{ij}^+)
\end{array} \right| & \mbox{for } i \neq j

\end{array} \right.
\label{F}
\end{equation}

\noindent
where $X_i$ is defined by (\ref{mag}) and for $U_{ij}$ we suppose
$U_{ij}=U_{ji}^+$. So $F$ is a sparse real $2N \times 2N$-matrix where
$N$ is number of sites.

\subsection{Partial solution of the MA gauge constraints.}

Having in mind a hybrid scheme \cite{Duane} to guide a MC simulation of
theory with partition function (\ref{zma}) we now define a set of independent,
with respect to constraints (\ref{mag}), variables $\{q\}$.

Let a field configuration $C_{MA} = \{U\}$ be a solution of (\ref{mag}). We
consider three link variables $S=\{U_{si}, U_{ij}, U_{jk}\}$ of the
configuration $C_{MA}$ such that links $L_S = \{(s,i),~(i,j),~(j,k)\}$ form a
continual path ($3l$-path) on the lattice.

Equations (\ref{mag}), which include link variables of set $S$, read

\begin{eqnarray}
\label{macon}
(U_{si}^q \sigma_3 U_{si}^{q+})^\bot
& = & (U_{si} \sigma_3 U_{si}^+)^\bot \nonumber \\
(U_{si}^{q+} \sigma_3 U_{si}^{q} ~+~ U_{ij}^q \sigma_3 U_{ij}^{q+})^\bot &
= & (U_{si}^+ \sigma_3 U_{si} ~+~ U_{ij} \sigma_3 U_{ij}^{+})^\bot \\
(U_{ij}^{q+} \sigma_3 U_{ij}^{q} ~+~ U_{jk}^q \sigma_3 U_{jk}^{q+})^\bot &
= & (U_{ij}^+ \sigma_3 U_{ij} ~+~ U_{jk} \sigma_3 U_{jk}^{+})^\bot
\nonumber \\
(U_{jk}^{q+} \sigma_3 U_{jk}^{q})^\bot & = & (U_{jk}^+ \sigma_3 U_{jk})^\bot
\nonumber
\end{eqnarray}

\noindent
where the terms independent on $S$ are omitted. Substitution $U_{si}^q =
U_{si}*\exp {(-{\it i} \sigma_3 q)}$  and  $U_{jk}^v = \exp {({\it i} \sigma_3
v)}*U_{jk}$ solves the first and last of equations (\ref{macon}). Two others are
equivalent to $4$ algebraic equations with $4$ independent variables, say
$u_{1,ij}, u_{2,ij}, u_{3,ij}$ and $v$, and allow to find a single solution
$U_{ij}^q$ and $v(q)$. So, for any $3l$-set $S$ we can locally define a new set
of variables

\begin{equation}
\vec{Q}_S = \{q,~Q_1^{\bot},~Q_2^{\bot},~Q_3^{\bot},~Q_4^{\bot}\}_S
= \{q,~\vec{Q}^{\bot}\}_S
\label{lv}
\end{equation}

\noindent
where $Q_i^{\bot}$ is the right-hand side of the $i$-th equation (\ref{macon}).
If the link variables apart from S are kept fixed, then the gauge-fixing
conditions require $\vec{Q}^{\bot}$ to be constant, whereas $q$ is a free
parameter.

Choosing a set of $3l$-paths $\{S\}$ such that $S_i \cap S_j = \emptyset
\mbox{ for } i \neq j$ we can locally define a set of independent variables
$\{q\}$. It allows to make transition $C_{MA} \rightarrow \tilde{C}_{MA}$
where the both configurations lay on the surface defined by the gauge
constraints (\ref{mag}).

\subsection{Hybrid Monte Carlo method for SU(2) gauge theory under the MA
gauge.}

By following the well known procedure we can eliminate the determinant in (\ref{zma})
and write the partition function in the form

\begin{equation}
Z=\int \delta(X^\bot) \left[d\mu (U) d(\Phi )\right] \exp (- S_W\{U\} -
\Phi^* F^{-1} \Phi )
\label{zphi}
\end{equation}

\noindent
where $\Phi$ is an additional complex scalar field. It looks very similar to
a partition function of gauge theory with pseudefermions except the
$\delta$-functions and form of the interaction matrix for the scalar field.

For numerical simulation of the system a Markov process should be constructed
with the fixed-point distribution defined by (\ref{zphi}), and the only
requirements for the transition probability of the process are detailed
balance and ergodicity.

The algorithm suggested in this paper includes two steps. First, we choose a
set $\{S\}$ of $N_{site}$ nonintersecting $3l$-paths on the lattice. The
choice is made at random from a uniform distribution. Second, we generate
a new field configuration by hybrid Monte Carlo method \cite{Duane}, so the
detailed balance and the gauge-fixing conditions are satisfied. We also
repeat the transition for different $\{S\}_i$ to gain ergodicity.

To specify the second step of the algorithm, let us take a configuration
$C_{MA}\{U\}$ satisfying the constrains (\ref{mag}) and select a set
$\{S\}$ of nonintersecting $3l$-paths on the lattice. Restricted to the link
variables associated with $\{S\}$, the desired probability distribution reads

$$
\left[ \delta(\vec{Q}^{\bot}-\vec{Q}_0^{\bot}) \left|
\Delta^{-1} (\vec{Q}, U) \right| d\vec{Q} d\Phi \right] \exp(-S_W\{U(q)\} -
\Phi^*F^{-1}\Phi )
$$

\noindent
where $\vec{Q}_0 \equiv \left. \vec{Q} \right|_{C_{MA}}$ and Jacobian
$\Delta$ is given by

$$
\Delta(\vec{Q},U_S) = Tr(\sigma_3 U_{ij} \sigma_3 U_{ij}^+)
Tr[\sigma_3(U_{ij} \sigma_3 U_{ij}^{+})(U_{ik} \sigma_3 U_{ik}^{+})]
$$

After integration over $\vec{Q}^{\bot}$ the probability distribution takes
the form

\begin{equation}
\left[ dq d\Phi \right] \exp(-S_W\{U(q)\} - \Phi^*F^{-1}\Phi - \sum_{i} \ln
(\left| \Delta(\vec{Q}, U) \right|)) = \left[ dq d\Phi \right] \exp(-S_{eff})
\label{qrho}
\end{equation}

\noindent
where the sum in the exponent is taken over all $3l$-paths of set $\{S\}$.

We apply the hybrid Monte Carlo algorithm \cite{Duane} with evolution in
pseudotime defined by $S_{eff}$ to generate a new MA configuration
$\tilde{C}_{MA}$.

It is easy to see that Markov process converging to (\ref{qrho}) in variables
$\vec{Q}$ yields a correct simulation of system (\ref{zphi}). Moreover, if
step-size in the integration of the equations of motion is taken small, a
quasicontinuous evolution in Langevin time keeps the configuration within
the Gribov horizon \cite{Zwanz}.

\section{Numerical simulation.}

The algorithm has been implemented for simulation of $SU(2)$ gauge theory on
the $4^4$ lattice. We started with a configuration very close to zero
field because matrix $F$ for the completely ordered configuration has zero
eigenvalues. The usual relaxation algorithm has been employed for initial MA
gauge fixing.

For the described in the previous section hybrid Monte Carlo algorithm, we
chose duration $\delta \tau = 0.0005$ for Langevin time steps and 400 steps
for every run. The choice was not optimized but yielded acceptance rate $\sim
0.7$. The generated configurations were separated by 12 Langevin runs with
different (random) sets of 256 $3l$-paths. It approximately reproduces a real
number of degrees of freedom of the system.

The first 500 configurations were skipped for thermolisation at $\beta$ values
4.0, ~3.0, ~2.5, ~2.3 and $\simeq 2000$ configurations at $\beta = 2.25$.
The next 100 configurations were used to calculate average plaquette. Every
configuration was searched for abelian monopoles \cite{Tous} but no monopoles
were found.

The results are summarized in Table 1. Here we also presented data on a
one-loop calculations of the plaquettes \cite{Karsch} for $8^4$ lattice and
values of the plaquette calculated by the heatbath algorithm. From the
collected data we could observe that the calculations under the MA gauge are
in an excellent agreement with the one-loop perturbative results at all $\beta$,
whereas the heatbath data essentially differ at $\beta \leq 2.5$.

\begin{center}
\samepage
Table 1. \\
Average plaquette in MA without monopoles, 1-loop expantion and full $SU(2)$.\\
\nopagebreak
\vspace{0.5cm}
\begin{tabular}{|lcccc|} \cline{1-5}
$\beta$ & $N_{mon}$ & $<Pl_{MA}>$  & $Pl_{one-loop}$ & $<Pl_{HB}>$ \\
\cline{1-5}
2.25 & 0 & 0.642(2) & 0.637 & 0.593 \\
2.3  & 0 & 0.647(3) & 0.645 & 0.609 \\
2.5  & 0 & 0.672(2) & 0.676 & 0.655 \\
3.0  & 0 & 0.729(3) & 0.733 & 0.724 \\
4.0  & 0 & 0.799(2) & 0.803 & 0.800 \\
\cline{1-5}
\end{tabular}
\end{center}

We additionally tested the stability of the generated configurations with
respect to a small random gauge transformation with a further gauge fixing by
the relaxation algorithm. In all cases the configurations appear to be
stable.  The situation changes crucially if matrix $F$ is replaced by
$\tilde{F}$ with $\tilde{f}_{ii} = f_{ii} \mbox{ and } \tilde{f}_{ij} = 0
\mbox{ for } i \neq j$. At $\beta = 2.25$ after $\sim 200$ sweeps starting
from an almost ordered MA configuration we receive a nonstable one sometimes
having abelian monopoles. So we come to the conclusion that it is the
Faddev--Popov determinant that keeps us within the Gribov horizon.

\section{Conclusion}

We have presented an algorithm for the numerical simulation of $SU(2)$ gauge
theory under the MA gauge. The first numerical results give evidence that
Gribov region for the MA gauge projection is split into at least two
separated sectors. The configurations laying within the locus of the first
zeros of $det(F)$ have no abelian monopoles.  Without monopoles the smallest
Wilson loop - plaquette becomes very close to its one-loop perturbative
value. It supports the idea that an effective theory could be built with
nonperturbative interaction of the residual abelian field with monopoles and
the other interactions are perturbative.

This work is supported in part by ISF Grant of Long Term Research Program
(No. RFW000).

\newpage

\end{document}